\documentstyle[prl,multicol,aps,psfig]{revtex}

\newcommand{\beq}{\begin{equation}}
\newcommand{\eeq}{\end{equation}}

\newcommand{\PP}{{\mathcal{P}}}
\newcommand{\NN}{{\mathcal{N}}}

\newcommand{\RR}{{\bf R}}
\newcommand{\RRp}{{\bf R'}}

\newcommand{\rr}{{\bf r}}

\newcommand{\mm}{{\bf m}}

\newcommand{\HH}{{\mathcal{H}}}

\newcommand{\BE}{\begin{equation}}
\newcommand{\EE}{\end{equation}}
\newcommand{\BEN}{\begin{eqnarray}}
\newcommand{\EEN}{\end{eqnarray}}

\begin{document}
\title{Path Integral Monte Carlo Calculation of the Deuterium Hugoniot}
\author{B. Militzer and D. M. Ceperley}
\address{Department of Physics \\
National Center for Supercomputing Applications \\
University of Illinois at Urbana-Champaign, Urbana, IL 61801}
\date{\today }
\maketitle

\begin{abstract}
Restricted path integral Monte Carlo simulations have been used to
calculate the equilibrium properties of deuterium for two
densities: $0.674$ and $0.838\,\rm{gcm}^{-3}$ ($r_s=2.00$ and
$1.86$) in the temperature range of $10\,000\,\rm{K} \le T \le
1\,000\,000\,\rm{K}$. Using the calculated internal energies and
pressures we estimate the shock hugoniot and compare with recent
Laser shock wave experiments. We study finite size effects and the
dependence on the time step of the path integral. Further, we
compare the results obtained with a free particle nodal
restriction with those from a self-consistent variational
principle, which includes interactions and bound states.
\end{abstract}

\vspace*{3mm}

\begin{multicols}{2}
\narrowtext

\section{Introduction}
Recent laser shock wave experiments on pre-compressed liquid
deuterium \cite{Si97,Co98} have produced an unexpected equation of
state for pressures up to 3.4 Mbar. It was found that deuterium
has a significantly higher compressibility than predicted by the
semi-empirical equation of state based on plasma many-body theory
and lower pressure shock data (see SESAME model \cite{Ke83}).
These experiments have triggered theoretical efforts to understand
the state of compressed hydrogen in this range of density and
temperature, made difficult because the experiments are in regime
where strong correlations and a significant degree of electron
degeneracy are present. At this high density, it is problematic
even to define the basic units such as molecules, atoms, free
deuterons and electrons. Conductivity measurements
\cite{Co98} as well as theoretical estimates \cite{Ro98,Mi99}
suggest that in the experiment, a state of significant but not
complete metalization was reached.

A variety of simulation techniques and analytical models have been
advanced to describe hydrogen in this particular regime. There are
{\em ab initio} methods such as restricted path integral Monte
Carlo simulations (PIMC) \cite{PC94,Ma96,Mi99} and density
functional theory molecular dynamics (DFT-MD) \cite{Le00,Ga99}. Further
there are models that minimize an approximate free energy function
constructed from known theoretical limits with respect to the
chemical composition, which work very well in certain regimes. The most
widely used include \cite{Sa95,ER85b,Ro98}.

We present new results from PIMC simulations. What emerges is a
relative consensus of theoretical calculations. First, we
performed a finite size and time step study using a parallelized
PIMC code that allowed simulation of systems with $N_P=64$ pairs of
electrons and deuterons and more importantly to decrease the time
step from $\tau^{-1}=10^6\,\rm{K}$ to $\tau^{-1}=8 \cdot
10^6\,\rm{K}$. More importantly, we studied the effect of the nodal
restriction on the hugoniot.

\section{Restricted path integrals}
The density matrix of a quantum system at temperature $k_BT=1 /
\beta $ can be written as a integral over all paths $\RR_t$, 
\beq
\rho(\RR_0 ,\RR_\beta;\beta) = \frac{1}{N!} \sum_\PP \: (\pm
1)^\PP \! \! \! \! \! \oint\limits_{\RR_0 \rightarrow \PP
\RR_{\beta}} \! \! \! \! \! \! d\RR_t \;\; e^{-S[\RR_t] }. 
\eeq
$\RR_t$ stands for the entire paths of $N$ particles in $3$
dimensional space $\RR_t = (\rr_{1t},\ldots,\rr_{Nt})$ beginning
at $\RR_0$ and connecting to $\PP \RR_\beta$. $\PP$ labels the
permutation of the particles. The upper sign corresponds to a
system of bosons and the lower one to fermions. For
non-relativistic particles interacting with a potential $V(\RR)$,
the action of the path $S[\RR_t]$ is given by, 
\beq S[\RR_t] =
\int_0^\beta \! dt \left[ \frac{m}{2} \left | \frac{d\RR(t)}{\hbar
dt} \right |^2+ V(\RR(t)) \right] + \mbox{const}. 
\eeq 
One can
estimate quantum mechanical expectation values using Monte Carlo
simulations \cite{Ce95} with a finite number of imaginary time
slices $M$ corresponding to a {\em time step} $\tau=\beta/M$.

For fermionic systems the integration is complicated due to the
cancellation of positive and negative contributions to the
integral, ({\em the fermion sign problem}). It can be shown that
the efficiency of the straightforward implementation scales like
$e^{-2 \beta N f}$, where $f$ is the free energy difference per
particle of a corresponding fermionic and bosonic system
\cite{Ce96}. In \cite{Ce91,Ce96}, it has been shown that one can
evaluate the path integral by restricting the path to only
specific positive contributions. One introduces a reference point 
$\RR^*$ on the path that specifies the nodes of the density matrix,
$\rho(\RR,\RR^*,t)=0$. A {\em node-avoiding} path for 
$ 0 < t \leq \beta $  neither touches nor crosses a node: 
$\rho(\RR(t),\RR^*,t) \not= 0$. 
By restricting the integral to node-avoiding paths, 
\BEN
\label{RPI} \nonumber \rho_F(\RR_{\beta} ,\RR^*;\beta) &=&\\ \int
\!\! d\RR_0 \: &\rho_F&(\RR_0, \RR^* ; 0) 
\! \! \! \! \! \! \! \! \! \! \! 
\oint\limits_{\RR_0 \rightarrow \RR_{\beta}\in \Upsilon(\RR^*)} 
\! \! \! \! \! \! \! \! \! \! \! \! \! 
d\RR_t \;\; e^{-S[\RR_t] }, 
\EEN 
($\Upsilon (\RR^*)$ denotes the restriction)
the contributions are positive and therefore PIMC represents, in
principle, a solution to the sign problem. The method is exact if
the exact fermionic density matrix is used for the restriction.
However, the exact density matrix is only known in a few cases. In
practice, applications have approximated the fermionic density
matrix, by a determinant of single particle density matrices, 
\beq
\rho(\RR,\RRp;\beta)=\left|
\begin{array}{ccc}
\rho_1(\rr_{1},\rr'_{1};\beta)&\ldots&\rho_1(\rr_{N},\rr'_{1};\beta)\\
\ldots&\ldots&\ldots\\
\rho_1(\rr_{1},\rr'_{N};\beta)&\ldots&\rho_1(\rr_{N},\rr'_{N};\beta)
\end{array}\right|
\quad.
\label{matrixansatz}
\eeq 
This approach has been extensively applied using the free
particle nodes\cite{Ce96},
\beq
 \label{freegauss}
 \rho_1(\rr,\rr',\beta) = (4 \pi \lambda \beta)^{-3/2} \: \mbox{exp}
 \left\{ -(\rr-\rr')^2/4 \lambda \beta \right\}
\eeq 
with $\lambda = \hbar^2/2m$, including applications to
dense hydrogen \cite{PC94,Ma96,Mi99}. It can be shown that for
temperatures larger than the Fermi energy the interacting nodal
surface approaches the free particle (FP) nodal surface. In addition, in
the limit of low density, exchange effects are negligible, the
nodal constraint has a small effect on the path and therefore its
precise shape is not important. The FP nodes also
become exact in the limit of high density when kinetic effects
dominate over the interaction potential.
However, for the densities and temperatures under consideration,
interactions could have a significant effect on the fermionic
density matrix.

To gain some quantitative estimate of the possible effect of the
nodal restriction on the thermodynamic properties, it is necessary
to try an alternative. In addition to FP nodes, we used
a restriction taken from a variational density matrix (VDM) that
already includes interactions and atomic and molecular bound states.

The VDM is a variational solution of the Bloch equation. Assume a
trial density matrix with parameters $q_i$ that depend on
imaginary time $\beta$ and $\RRp$, 
\beq
\rho(\RR,\RRp;\beta)=\rho(\RR,q_1,\ldots,q_m)
\;.
\eeq
By minimizing the integral: \beq \int \! d\RR \left(
\frac{\partial \rho(\RR,\RRp;\beta)}{\partial \beta}+\HH \,
\rho(\RR,\RRp;\beta) \right)^{\!\!2} =0\quad, \label{varprinc}
\eeq 
one determines equations  for the dynamics of the parameters
in imaginary time:
\beq
   \label{matrix}
    \frac{1}{2}\frac{\partial H}{\partial \vec{q}}\; + \;\,
    \stackrel{{\textstyle \leftrightarrow}}{\NN}\: \dot{\vec{q}}=  0
    \;\;\;\;\mbox{where}\;\;\;\;
    H \equiv \int \rho \HH \rho\;d\RR  \;.
\label{H}
\eeq 
The normalization matrix is:
\BEN \NN_{ij} &=&
\lim_{q'\rightarrow q} \frac{\partial^{\,2}}{ \partial q_i
\partial q'_j}\left[\int \!d\RR \;
\rho(\RR,\vec{q}\,;\beta) \; \rho(\RR,\vec{q}\:'\,;\beta)\right]
\;.
\label{NN} 
\EEN
We assume the density matrix is a Slater determinant of single
particle Gaussian functions
\beq
 \label{gauss}
 \rho_1(\rr,\rr',\beta) = (\pi w)^{-3/2} \: \mbox{exp}
 \left\{ -(\rr-\mm)^2/w + d \right\}
\eeq 
where the variational parameters are the mean $\mm$,
squared width $w$ and amplitude $d$. The differential equation for
this ansatz are given in \cite{MP00}. The initial conditions at
$\beta\longrightarrow 0$ are $w= 2\beta$, $\mm=\rr'$ and $d=0$ in
order to regain the correct FP limit.
It follows from Eq.~\ref{varprinc} that at low temperature, the
VDM goes to the lowest energy wave function within the variational
basis. For an isolated atom or molecule this will be a bound
state, in contrast to the delocalized state of the FP
density matrix. A further discussion of the VDM properties is
given in \cite{MP00}. 
Note that this discussion concerns only the nodal
restriction. In performing the PIMC simulation, the complete
potential between the interacting charges is taken into account
as discussed in detail in \cite{Ce95}. 

\begin{figure}[htb]
\centerline{\psfig{figure=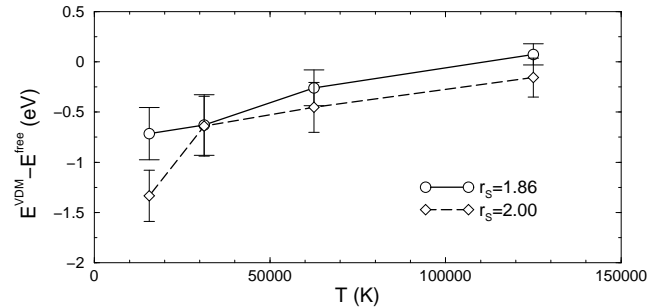,width=84mm,angle=270}}
\caption{Difference in the internal energy from PIMC simulations
with VDM and FP nodes vs. temperature using $N_P=32$ and 
$\tau^{-1}=2 \cdot 10^6 \rm{K}$.}
\label{energydiff}
\end{figure}

Simulations with VDM nodes lead to lower internal energies than
those with FP nodes as shown in Fig. \ref{energydiff}.
Since the free energy $F$ is the integral of the internal energy
over temperature, one can conclude that VDM nodes yield to a smaller $F$
and hence, are the more appropriate nodal surface.

For the two densities considered here, the state of deuterium goes
from a plasma of strongly interacting but un-bound deuterons and electrons
at high $T$ to a regime at low $T$, which is characterized by a significant
electronic degeneracy and bound states. Also at decreasing $T$, one finds an 
increasing number of electrons involved in long permutation cycles. Additionally
for $T \le 15\,625\,\rm{K}$, molecular formation is observed. Comparing
FP and VDM nodes, one finds that VDM predicts a higher molecular fraction
and fewer permutations hinting to more localized electrons.

\section{Shock Hugoniot}

The recent experiments measured the shock velocity, propagating
through a sample of pre-compressed liquid deuterium characterized
by an initial state, ($E_0$,~$V_0$,~$p_0$) with $T=19.6\,\rm{K}$
and $\rho_0=0.171\,\rm{g/cm^3}$. Assuming an ideal planar shock
front, the variables of the shocked material ($E$,~$V$,~$p$)
satisfy the hugoniot relation \cite{Ze66},
\begin{equation}
H = E-E_0+\frac{1}{2}(V-V_0)(p+p_0)=0 \quad.
\end{equation}
We set $E_0$ to its exact value of $-15.886\rm{eV}$ per atom
\cite{KW64} and $p_0 = 0$. Using the simulation results for $p$
and $E$, we calculate $H(T,\rho)$ and then interpolate $H$
linearly at constant $T$ between the two densities corresponding
to $r_s = 1.86$ and $2$ to obtain a point on the hugoniot in the
$(p,\rho)$ plane. (Results at $r_s = 1.93$ confirm the function is
linear within the statistical errors). The PIMC data for $p$, $E$, and
the hugoniot are given in Tab. \ref{table1}.

\begin{figure}[htb]
\centerline{\psfig{figure=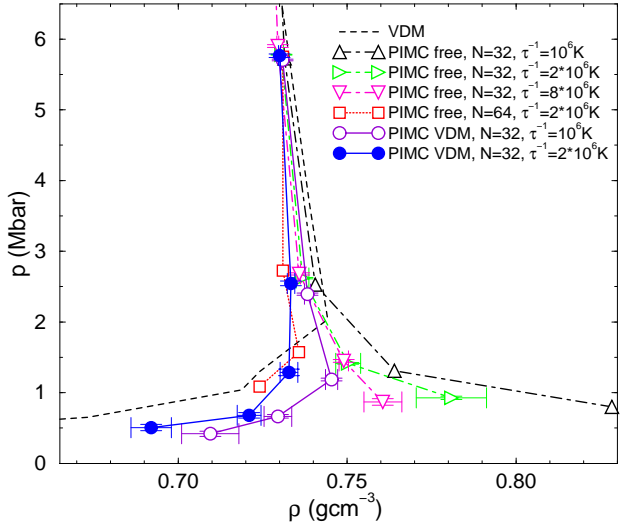,width=82mm,angle=270}}
\caption{Comparison hugoniot function calculated with PIMC
simulations of different accuracy: FP nodes 
with $N_P$=32 ($\triangle$ for $\tau^{-1}=10^6\rm K$ reported in [5]
, $\rhd$ for $\tau^{-1}=2 \cdot 10^6 \rm K$, $\bigtriangledown$ 
for $\tau_F^{-1}=8 \cdot 10^6 \rm{K}$ and
$\tau_B^{-1}=2 \cdot 10^6 \rm{K}$) and $N_P$=64 
($\Box$ for $\tau^{-1}=2 \cdot 10^6 \rm{K}$) as well as with 
VDM nodes and $N_P$=32 ($\circ$ for $\tau^{-1}=10^6\rm{K}$ and
$\bullet$ for $\tau^{-1}=2 \cdot 10^6\rm{K}$). Beginning at high pressures,
the points on each hugoniot correspond to the following
temperatures $125\,000, 62\,500, 31\,250, 15\,625,$ and
$10\,000\,\rm{K}$. The dashed line corresponds to a calculation using
the VDM alone.}
\label{hugoniotFigure1}
\end{figure}

In Fig. \ref{hugoniotFigure1}, we compare the effects of different
approximations made in the PIMC simulations such as time step
$\tau$, number of pairs $N_P$ and the type of nodal restriction. For
pressures above 3 Mbar, all these approximations have a very small
effect. The reason is that PIMC simulation become increasingly
accurate as temperature increases. The first noticeable difference
occurs at $p \approx 2.7 \rm{Mbar}$, which corresponds to
$T=62\,500\,\rm{K}$. At lower pressures, the differences become
more and more pronounced. We have performed simulations with free
particle nodes and $N_P=32$ for three different values of $\tau$. 
Using a smaller time step
makes the simulations computationally more demanding and it
shifts the hugoniot curves to lower densities. These differences
come mainly from enforcing the nodal surfaces more accurately,
which seems to be more relevant than the simultaneous improvements
in the accuracy of the action $S$, that is the time step is
constrained more by the Fermi statistics than it is by the
potential energy. 
We improved the efficiency of the algorithm by using a smaller time
step $\tau_F$ for evaluating the Fermi action than the time step
$\tau_B$ used for the potential action. Unless specified otherwise,
we used $\tau_F=\tau_B=\tau$.
At even lower pressures not shown in Fig.
\ref{hugoniotFigure1}, all of the hugoniot curves with FP nodes 
turn around and go to low densities as expected.

As a next step, we replaced the FP nodes by VDM nodes.
Those results show that the form
of the nodes has a significant effect for $p$ below 2 Mbar. Using
a smaller $\tau$ also shifts the curve to slightly lower
densities. In the region where atoms and molecules are forming, it
is plausible that VDM nodes are more accurate than free nodes
because they can describe those states \cite{MP00}. We also show a
hugoniot derived on the basis of the VDM alone (dashed line).
These results are quite reasonable considering the approximations
(Hartree-Fock) made in that calculation. Therefore, we consider
the PIMC simulation with the smallest time step using VDM nodes
($\bullet$) to be our most reliable hugoniot. Going to bigger system
sizes $N_P=64$ and using FP nodes also shows a shift towards lower
densities.

\begin{figure}[htb]
\centerline{\psfig{figure=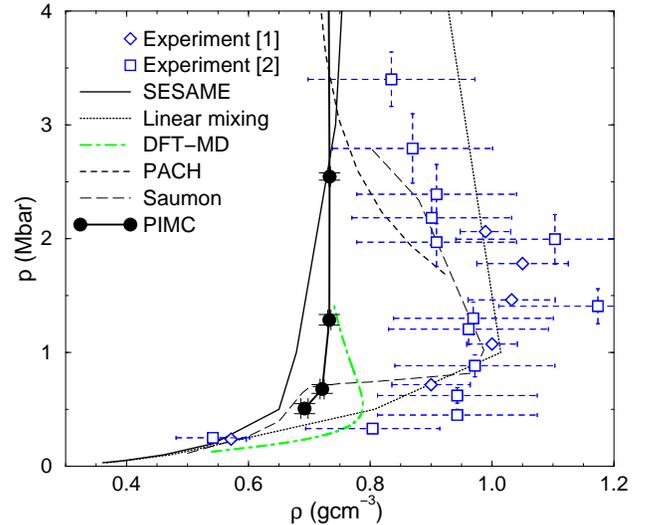,width=82mm,angle=270}}
\caption{ Comparison of experimental and several theoretical
Hugoniot functions.  The PIMC curve was calculated with VDM
nodes, $\tau^{-1}=2 \cdot 10^6\,\rm{K}$, and $32$ pairs of
electrons and deuterons.} \label{hugoniotFigure2}
\end{figure}

Fig. \ref{hugoniotFigure2} compares the Hugoniot from Laser shock
wave experiments \cite{Si97,Co98} with PIMC simulation (VDM nodes,
$\tau^{-1}=2 \cdot 10^6\,\rm{K}$) and several theoretical
approaches: SESAME model by Kerley \cite{Ke83} (thin solid
line), linear mixing model by Ross (dashed line) \cite{Ro98},
DFT-MD by Lenosky
{\it et al.} \cite{Le00} (dash-dotted line), Pad\'e approximation
in the chemical picture (PACH) by Ebeling {\it et al.}
\cite{ER85b} (dotted line), and the work by Saumon {\it et al.}
\cite{Sa95} (thin dash-dotted line).

The differences of the various PIMC curves in Fig.
\ref{hugoniotFigure1} are small compared to the deviation from the
experimental results \cite{Si97,Co98}. There, an increased
compressibility with a maximum value of $6 \pm 1$ was found while
PIMC predicts $4.3 \pm 0.1$, only slightly higher than that given
by the SESAME model. Only for $p>2.5 \rm Mbar$, does our
hugoniot lie within experimental errorbars. In this regime, the
deviations in the PIMC and PACH hugoniot are relatively small, less
than $0.05 \, \rm gcm^{-3}$ in density. In the high pressure
limit, the hugoniot goes to the FP limit of 4-fold
compression. This trend is also present in the experimental
findings. For pressures below 1 Mbar, the PIMC hugoniot goes back
lower densities and shows the expected tendency towards the
experimental values from earlier gas gun work \cite{Ne83,Ho95} and
lowest data points from \cite{Si97,Co98}. For these low pressures,
differences between PIMC and DFT-MD are also relatively small.

\section{Conclusions}
We reported results from PIMC simulations and performed a finite
size and time step study. Special emphasis was put on improving
the fermion nodes where we presented the first PIMC results with
variational instead of FP nodes. We find a slightly
increased compressibility of $4.3 \pm 0.1$ compared to the SESAME
model but we cannot reproduce the experimental findings of
values of about $6 \pm 1$.  Further theoretical and experimental
work will be needed to resolve this discrepancy.

\acknowledgements The authors would like to thank W. Magro for the
collaboration concerning the parallel PIMC simulations and E.L. Pollock
for the contributions to the VDM method. This work was supported
by the CSAR program and the Department of Physics at the
University of Illinois. We used the computational facilities at
the National Center for Supercomputing Applications and Lawrence
Livermore National Laboratories.

\end{multicols}
\widetext

\begin{table}
\vspace*{-4.0mm}

\caption{Pressure $p$ and internal energy per atom $E$ from PIMC
simulations with $32$ pairs of electrons and deuterons. For $T \ge
250\,000\,\rm{K}$, we list results from simulations with FP nodes 
and $\tau_F^{-1}=8 \cdot 10^6\,\rm{K}$ and
$\tau_B^{-1}=2 \cdot 10^6\,\rm{K}$, otherwise with VDM nodes and
$\tau^{-1}=2 \cdot 10^6\,\rm{K}$.}
\begin{tabular}{ccccccccc}
$T (K)$ &
$p (Mbar), r_s=2 $ &
$E (\rm{eV}), r_s=2$ &
$p (\rm{Mbar}), r_s=1.86 $ &
$E (\rm{eV}), r_s=1.86$ &
$\rho^{Hug} (\rm{gcm}^{-3})$ &
$p^{Hug} (\rm{Mbar})$\\
\tableline
1000000 &53.79 $\pm$ 0.05  &  245.7 $\pm$ 0.3 & 66.85 $\pm$ 0.08 & 245.3 $\pm$ 0.4  & 0.7019 $\pm$ 0.0001 &56.08 $\pm$ 0.05\\
~500000 &25.98 $\pm$ 0.04  &  113.2 $\pm$ 0.2 & 32.13 $\pm$ 0.05 & 111.9 $\pm$ 0.2  & 0.7130 $\pm$ 0.0001 &27.48 $\pm$ 0.04\\
~250000 &12.12 $\pm$ 0.03  &  ~45.7 $\pm$ 0.2 & 14.91 $\pm$ 0.03 & ~44.3 $\pm$ 0.2  & 0.7242 $\pm$ 0.0001 &12.99 $\pm$ 0.02\\
~125000 &~5.29 $\pm$ 0.04  &  ~11.5 $\pm$ 0.2 & ~6.66 $\pm$ 0.02 & ~11.0 $\pm$ 0.1  & 0.7300 $\pm$ 0.0003 &~5.76 $\pm$ 0.02\\
~~62500 &~2.28 $\pm$ 0.04  &  ~-3.8 $\pm$ 0.2 & ~2.99 $\pm$ 0.04 & ~-3.8 $\pm$ 0.2  & 0.733  $\pm$ 0.001  &~2.54 $\pm$ 0.03\\
~~31250 &~1.11 $\pm$ 0.06  &  ~-9.9 $\pm$ 0.3 & ~1.58 $\pm$ 0.07 & ~-9.7 $\pm$ 0.3  & 0.733  $\pm$ 0.003  &~1.28 $\pm$ 0.05\\
~~15625 &~0.54 $\pm$ 0.05  &  -12.9 $\pm$ 0.3 & ~1.01 $\pm$ 0.05 & -12.0 $\pm$ 0.2  & 0.721  $\pm$ 0.004  &~0.68 $\pm$ 0.04\\
~~10000 &~0.47 $\pm$ 0.05  &  -13.6 $\pm$ 0.3 & ~0.80 $\pm$ 0.08 & -13.2 $\pm$ 0.4  & 0.690  $\pm$ 0.007  &~0.51 $\pm$ 0.05\\
\end{tabular}
\label{table1}
\end{table}
%
\vspace*{-5mm}
\begin{multicols}{2}

\end{multicols}


\begin{thebibliography}{10}

\bibitem{Si97}
I.~B. Da~Silva {\it et al.}
\newblock {\em Phys. Rev. Lett.}, {\bf 78}:783, 1997.

\bibitem{Co98}
G.~W.~Collins {\it et al.}
\newblock {\em Science}, {\bf 281}:1178, 1998.

\bibitem{Ke83}
G.~I. Kerley.
\newblock Molecular based study of fluids.
\newblock page 107. ACS, Washington DC, 1983.

\bibitem{Ro98}
M.~Ross.
\newblock {\em Phys. Rev. B}, {\bf 58}:669, 1998.

\bibitem{Mi99}
B.~Militzer, W.~Magro, and D.~Ceperley.
\newblock {\em Contr. Plasma Physics}, {\bf 39 }1-2:152, 1999.

\bibitem{PC94}
C.~Pierleoni, D.M. Ceperley, B.~Bernu, and W.R. Magro.
\newblock {\em Phys. Rev. Lett.}, {\bf 73}:2145, 1994.

\bibitem{Ma96}
W.~R. Magro, D.~M. Ceperley, C.~Pierleoni, and B.~Bernu.
\newblock {\em Phys. Rev. Lett.}, {\bf 76}:1240, 1996.

\bibitem{Le00}
T.~J. Lenosky, S.~R. Bickham, J.~D. Kress, and L.~A. Collins.
\newblock {\em Phys. Rev. B}, 61:1, 2000.

\bibitem{Ga99}
G.~Galli, R.Q. Hood, A.U. Hazi, and F.~Gygi.
\newblock {\em in press, Phys. Rev. B}, 1999.

\bibitem{Sa95}
D.~Saumon, G.~Chabrier, and H.~M.~Van Horn.
\newblock {\em Astrophys. J.}, 99 2:713, 1995.

\bibitem{ER85b}
W.~Ebeling and W.~Richert.
\newblock {\em Phys. Lett. A}, {\bf 108}:85, 1985.

\bibitem{Ce95}
D.~M. Ceperley.
\newblock {\em Rev. Mod. Phys.}, 67:279, 1995.

\bibitem{Ce96}
D.~M. Ceperley.
\newblock Monte carlo and molecular dynamics of condensed matter systems.
\newblock Editrice Compositori, Bologna, Italy, 1996.

\bibitem{Ce91}
D.~M. Ceperley.
\newblock {\em J. Stat. Phys.}, 63:1237, 1991.

\bibitem{MP00}
B.~Militzer and E.~L. Pollock.
\newblock {\em in press, Phys. Rev. E}, 2000.

\bibitem{Ze66}
Y.~B. Zeldovich and Y.~P. Raizer.
\newblock {\em Physics of Shock Waves and High-Temperature Hydrodynamic
  Phenomena}.
\newblock Academic Press, New York, 1966.

\bibitem{KW64}
W.~Kolos and L.~Wolniewicz.
\newblock {\em J. Chem. Phys.}, 41:3674, 1964.

\bibitem{Ne83}
W.J. Nellis and A.C.~Mitchell {\it et al.}
\newblock {\em J. Chem. Phys.}, 79:1480, 1983.

\bibitem{Ho95}
N.C.Holmes, M.~Ross, and W.J.Nellis.
\newblock {\em Phys. Rev. B}, 52:15835, 1995.

\end{thebibliography}
\end{document}